\documentclass[pdflatex,sn-basic,Numbered]{sn-jnl}% Basic Springer Nature Reference Style/Chemistry Reference Style
\usepackage{graphicx}%
\usepackage{multirow}%
\usepackage{amsmath,amssymb,amsfonts}%
\usepackage{amsthm}%
\usepackage{mathrsfs}%
\usepackage[title]{appendix}%
\usepackage{xcolor}%
\usepackage{textcomp}%
\usepackage{manyfoot}%
\usepackage{booktabs}%
\usepackage{algorithm}%
\usepackage{algorithmicx}%
\usepackage{algpseudocode}%
\usepackage{listings}%
\usepackage{xurl}

\usepackage[disable]{todonotes}%
\usepackage{tikz}
\usetikzlibrary{shapes.geometric, arrows, calc, fit, arrows.meta}
\usepackage{color,soul}
\usepackage{subcaption}
\usepackage[normalem]{ulem}

\theoremstyle{thmstyleone}%
\theoremstyle{thmstyletwo}%

\theoremstyle{thmstylethree}%

\unnumbered% uncomment this for unnumbered level heads

\begin{document}

\title[Frequentist and Bayesian population pharmacometrics]{\centering Frequentist and Bayesian population pharmacometrics: similarities, differences, and practical implications}

\author*[1]{\fnm{Jodie A.} \sur{Cochrane}}\email{jodie.cochrane@it.uu.se}
\author[2]{\fnm{Haini} \sur{Wen}}\email{haini.wen@uu.se}
\author[2,3]{\fnm{Liang} \sur{Yang}}\email{liang.yang@csu.edu.cn}
\author[4]{\fnm{Tyler} \sur{Cassidy}}\email{tyler.cassidy@ubc.ca}
\author[1]{\fnm{Sara} \sur{Hamis}}\email{sara.hamis@it.uu.se}

\affil*[1]{\orgdiv{Department of Information Technology}, \orgname{Uppsala University}, \orgaddress{\city{Uppsala}, \country{Sweden}}}
\affil[2]{\orgdiv{Department of Pharmacy}, \orgname{Uppsala University}, \orgaddress{\city{Uppsala}, \country{Sweden}}}
\affil[3]{\orgdiv{Xiangya School of Pharmaceutical Sciences}, \orgname{Central South University}, \orgaddress{\city{Changsha}, \country{China}}}
\affil[4]{\orgdiv{Faculty of Pharmaceutical Sciences}, \orgname{University of British Columbia}, \orgaddress{\city{Vancouver}, \country{Canada}}}

\abstract{
Although the frequentist versus Bayesian debate has a long-standing history, and both are well-established approaches for inference, they are often considered separately in pharmacometric literature, with their theoretical and practical differences receiving limited direct comparison.
We address this gap by providing a conceptual background and comparative evaluation of frequentist and Bayesian methods for parameter inference in pharmacometrics. 
First, we highlight the theoretical and algorithmic differences between approaches within a general and pharmacometric context.
We then apply both to classical compartmental pharmacokinetic models using simulated and real data; initially through evaluating four inference methods (two frequentist, two Bayesian) across a range of simulated data conditions, followed by investigating each inference approach in the practical setting. 
The simulation studies facilitate evaluation of the sensitivity of the methods to sample size, number of observations, residual noise and inter-individual variability, comparing the performance with respect to relative error, confidence/credible interval coverage and bias in the estimated variability. 
The real-world data analysis further highlights how each approach quantifies and presents uncertainty estimates by examining this for both the parameter estimates and system dynamics.
By providing a theoretical and practical comparison that highlights the similarities and differences between approaches, this work aims to lower the barrier to understanding and implementing Bayesian and frequentist inference in pharmacometrics.
}

\keywords{Bayesian inference, Pharmacokinetics, Compartmental analysis, Modelling}

\maketitle

\section{Introduction}\label{sec1}

Pharmacometric modelling plays a central role in modern drug development, informing decisions from early-phase dose selection to regulatory approval \cite{Madabushi2022}. 
This role is reinforced by active encouragement of quantitative model-informed drug development (MIDD) approaches from regulatory bodies worldwide, including the European Medicines Agency (EMA) \cite{EMA2018}, the U.S. Food and Drug Administration (FDA) \cite{FDA2022}, Japan's Pharmaceuticals and Medical Devices Agency (PMDA) \cite{Kijima2022}, and MIDD guideline alignment efforts by the International Council for Harmonisation (ICH) \cite{ICH2026}. 
As reliance on pharmacometric models grows, so too does the importance of how their parameters are estimated from data. For mixed-effects models, which are the gold standard in population pharmacometrics \cite{Owen2014}, and the focus of this paper, this means inferring both population-level (fixed-effect) and individual-level (random-effect) parameters. This task can be approached from two distinct statistical philosophies: frequentist and Bayesian inference. 

%
%  Freq Tools
%
Frequentist approaches to parameter estimation in pharmacometrics have a long history, with software tools dating back to the introduction of NONMEM for mixed-effects models in 1980 \cite{Csajka2006}.  
Today, several such tools are available, ranging from proprietary, closed-source platforms such as NONMEM \cite{Bauer2019}, to closed-source tools that are free for academic use, such as Monolix \cite{Traynard2020}, to fully free, open-source alternatives such as nlmixr2 \cite{fidler2019nonlinear}. 
Moreover, several software libraries have emerged that enable researchers to create customisable implementations of, e.g., pharmacometric algorithms in common programming languages like R \cite{Comets2017_R}, Python \cite{Chen2024_python}, and MATLAB \cite{Hosseini2018_matlab}.

%
%  Bay Tools
%
Bayesian approaches, by contrast, have a shorter track record in pharmacometrics. Although Bayesian pharmacometric modelling using Markov chain Monte Carlo (MCMC) methods, such as Gibbs sampling, dates back to the 1990s \cite{gelman1996physiological}, poor scalability and slow convergence on complex hierarchical nonlinear models initially limited its practical use. 
It was only with the development of Hamiltonian Monte Carlo (HMC) and its implementation in accessible software such as Stan \cite{Carpenter2017}, first released in 2012, that Bayesian inference became a computationally competitive option for pharmacometric model inference. 
This gives frequentist approaches a two-decade head start in practice, tooling, and regulatory familiarity. However, recent work highlights Bayesian pharmacometric approaches, now ranging from purpose-built tools such as Torsten \cite{margossian2022flexible} (built on Stan), to Bayesian estimation options within NONMEM itself (available since NONMEM 7 \cite{Johnston2024}), to tutorials on custom implementations in programming languages like Julia and Stan \cite{elmokadem2023bayesian}.

%
%  the freq vs Bay split
%
Frequentist and Bayesian approaches are typically treated as separate camps in the literature: pharmacometric tutorials and workflow studies commonly focus on either frequentist \cite{mould2013basic} or Bayesian \cite{elmokadem2023bayesian,margossian2022flexible} methods, and comparison papers typically compare methods within either approach \cite{Schoemaker2019,Tsiros2019}. 
While philosophically reasonable, this separation is somewhat artificial in practice, as popular mixed-effects optimisation methods such as FOCEI and SAEM are mainly labelled as frequentist but incorporate elements of Bayesian reasoning. 
This is most clearly illustrated by FOCEI, which, during optimisation, computes an empirical Bayes estimate (EBE) of each patient's individual-level parameters 
via a method that encourages shrinkage of the estimate toward the population typical value, acting as an implicit prior \cite{Savic2009}---priors being a concept inherent to Bayesian modelling, and one classically criticised by frequentists \cite{Efron1986}. 
Further, user demand to move more readily between frequentist and Bayesian pharmacometric modelling is reflected in NONMEM's own recent adoption of Bayesian estimation methods \cite{Johnston2024}.

%
%  our contribution
%
In this paper, we contribute to addressing this frequentist versus Bayesian dichotomy in pharmacometric modelling through a practical, side-by-side comparison of parameter inference methods for classical pharmacokinetic (PK) models. 
We highlight the practical advantages and disadvantages of each approach, and characterise their differences and similarities. 
We do this in three steps: first, by walking through the core theory, algorithms, and assumptions that underlie each approach; second, by applying both to simulated data, where known ground truth allows direct validation; and third, by applying them to a real dataset, demonstrating practical application.

\section{Background}\label{sec:background}

\subsection{Overview of parameter inference}

A central part of modelling is inference, in which we wish to find the parameters $\boldsymbol{\Theta}$ such that the model predictions best fit the observed data $\mathbf{Y}$. 
The model itself relates some inputs of the system to outputs, and in pharmacometrics is often guided by underlying physical processes. 
Both frequentist and Bayesian methods can be used to learn the model parameters.
The main differences between the two are how the parameters are inferred and reported. 

\subsubsection{Frequentist approach}\label{sec:freq}

Frequentist statistics considers the probability of an event to be based on the long-run frequency at which the event will occur. For inference, this approach assumes that the model parameters have fixed, unknown values, with randomness arising from the data-generating process.
This approach is often posed as an optimisation problem, where the function to be optimised is the likelihood of observing the data $\mathbf{Y}$ given the model parameters $\boldsymbol{\Theta}$, and is denoted by $p(\mathbf{Y}\mid\boldsymbol{\Theta})$.
The model parameters that best describe the data are obtained by maximising this likelihood function, with the resulting parameter values referred to as the maximum-likelihood (ML) estimate.
This is expressed mathematically as,
\begin{equation}\label{eq:ml}
    \hat{\boldsymbol{\Theta}}_{\text{ML}} = \text{arg}\max_{\boldsymbol{\Theta}} \enspace p(\mathbf{Y}\mid\boldsymbol{\Theta}),
\end{equation}
where $\hat{\boldsymbol{\Theta}}_{\text{ML}}$ denotes the parameter values corresponding to the ML estimate.

Solving this optimisation problem produces a single point-estimate value for each parameter. However, such a point estimate is based on treating the observed data as fixed. 
In reality, this is not the case, as repeated experiments or additional individuals would likely exhibit different (but similar) observational data. The uncertainty in the data itself should thus correspond to an uncertainty in the estimated parameter values. Frequentists use confidence intervals to express this uncertainty.
These intervals define a range of values where, under repeated experiments, a specified proportion (the confidence level) of intervals will contain the true parameter value.
As such, frequentist methods typically report both point estimates and corresponding confidence intervals based on the assumption that the parameters are fixed, but the data are inherently random.

\subsubsection{Bayesian approach}\label{sec:bayes}

Bayesian, in contrast to frequentist, statistics assumes that all quantities not known with certainty are treated as random and modelled using probability distributions, including the model parameters themselves. 
The fundamental equation is Bayes' Rule,
\begin{equation}\label{eq:bayes}
    p(\boldsymbol{\Theta}\mid\mathbf{Y}) = \frac{p(\mathbf{Y}\mid\boldsymbol{\Theta})p(\boldsymbol{\Theta})}{p(\mathbf{Y})} .
\end{equation}
The term on the left-hand side represents the posterior distribution $p(\boldsymbol{\Theta}\mid\mathbf{Y})$, which is the quantity of interest. Computing this requires two definitions. The first is the likelihood model $p(\mathbf{Y}\mid\boldsymbol{\Theta})$, which is equivalent to that used in the frequentist approach, and defines the likelihood of observing the data given the current set of model parameters. 
The second is the prior distribution, denoted $p(\boldsymbol{\Theta})$, which incorporates any existing knowledge about the parameters. 
The denominator $p(\mathbf{Y})$, known as the evidence, acts as a normalising constant, and does not necessarily need to be computed in order to evaluate the posterior. 

Uncertainty estimates in the Bayesian setting, in this case called credible intervals, can be computed from the posterior distribution and provide an interval containing the parameter with a specified posterior probability. Point estimates, if desired, are computed similarly to the frequentist case, but here the posterior distribution is instead maximised to give the maximum a posteriori (MAP) estimate, 
\begin{equation}\label{eq:map}
    \hat{\boldsymbol{\Theta}}_{\text{MAP}} = \text{arg}\max_{\boldsymbol{\Theta}} \enspace p(\boldsymbol{\Theta}\mid\mathbf{Y}).
\end{equation}
Note that here we are choosing the most likely parameters given the data, in contrast to the ML setting (Eq.~\ref{eq:ml}), in which we choose the parameters that make the data most likely.

\subsubsection{How do frequentist and Bayesian approaches relate?}
\begin{figure}[t]
    \centering
    \includegraphics[width=\linewidth]{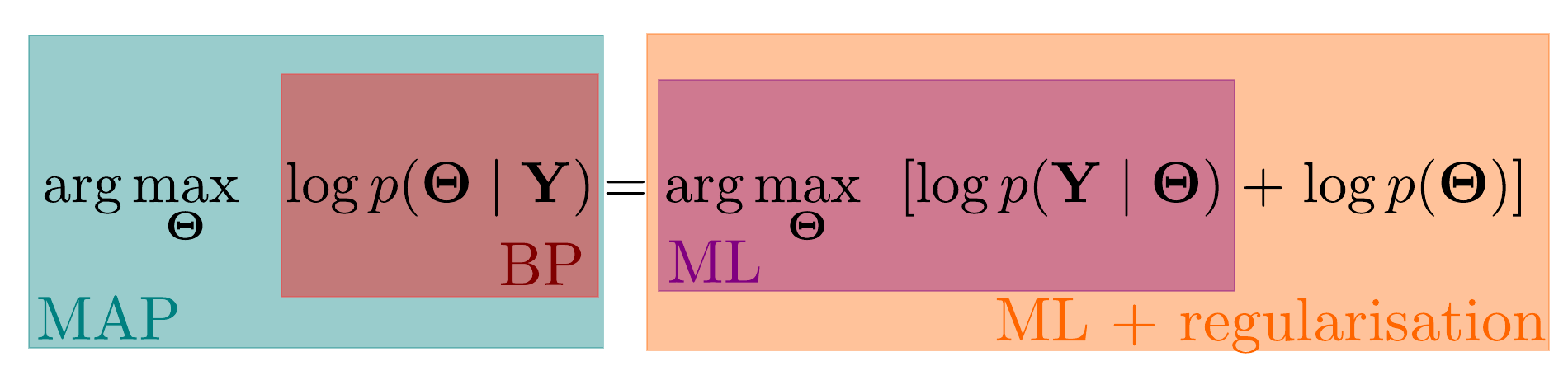}
    \caption{
    The relationship between frequentist and Bayesian inference approaches.
    Here, $\boldsymbol{\Theta}$ denotes the model parameters and $\mathbf{Y}$ the observed data.
    Frequentist algorithms primarily focus on the maximum likelihood (ML) solution, as shown in the inner right box (purple), which can be extended to include regularisation through the outer right box (orange). The maximum a posteriori (MAP) estimate, a concept from Bayesian statistics, in fact aligns with the regularised ML estimate and is shown in the outer left box (blue). This is computed from the Bayesian posterior (BP) that is the target of such inference methods, as shown in the inner left box (red) }
    \label{fig:comp_bayes_freq}
\end{figure}

Although frequentist and Bayesian inference approaches differ in their underlying philosophies, the relationship in Eq.~\ref{eq:bayes} reveals a connection between them. By taking the logarithm of both sides of Eq.~\ref{eq:bayes}, we get,
\begin{equation}\label{eq:bayes-log}
    \log p(\boldsymbol{\Theta}\mid\mathbf{Y}) = \log p(\mathbf{Y}\mid\boldsymbol{\Theta}) + \log p(\boldsymbol{\Theta}) - \log p(\mathbf{Y}).
\end{equation}
Optimising this equation with respect to the model parameters $\boldsymbol{\Theta}$ gives,
\begin{equation}\label{eq:map-ml}
    \text{arg}\max_{\boldsymbol{\Theta}} \enspace \log p(\boldsymbol{\Theta}\mid\mathbf{Y}) = 
    \text{arg}\max_{\boldsymbol{\Theta}} \enspace \left[ \log p(\mathbf{Y}\mid\boldsymbol{\Theta})\right. + 
    \left.\log p(\boldsymbol{\Theta})\right], 
\end{equation}
where the last term in Eq.~\ref{eq:bayes-log} is omitted as it is constant with respect to $\boldsymbol{\Theta}$.
Comparing back to Eqs. \ref{eq:ml} and \ref{eq:map}, we notice that the expressions for both the MAP and ML appear (note that this is a well-established connection, see \cite{park2008bayesian}). Figure \ref{fig:comp_bayes_freq} gives a visual representation of this connection between Bayesian approaches (target posterior and MAP estimate) and frequentist approaches (ML estimates with and without regularisation). The additional term $ \log p(\boldsymbol{\Theta})$ that connects the two represents the prior distribution in log form and, when used within a machine learning context, is referred to as regularisation. We also see the connection to the posterior distribution $p(\boldsymbol{\Theta}\mid\mathbf{Y})$, the target of Bayesian methods. 

\subsection{Inference in pharmacometrics}

We now apply the inference methods described above to pharmacometrics. In many pharmacometric applications, we wish to model the movement of a drug through a patient. To then infer the model parameters, we often maximise the likelihood, or equivalently the log-likelihood, to find the most likely parameters given the data. In general, the likelihood can be written as:
\begin{equation}\label{eq:llh}
    L(\boldsymbol{\Theta}) := p(\mathbf{Y} \mid \boldsymbol{\Theta}) = \prod_i p(\boldsymbol{y}_i \mid \boldsymbol{\Theta}),
\end{equation}
where $\boldsymbol{y}_i = (y_{i,1}, \dots, y_{i,N_i})$ represents the $N_i$ observations for patient $i$, and the assumed independence between individuals results in a product representation.

%%%

In mixed-effects models, additional flexibility is included in the model to account for patient variability \cite{lindstrom1990nonlinear}. This is achieved by assuming that each individual has their own set of parameters that are a combination of population-level (fixed-effect) parameters $\boldsymbol{\theta}_{pop}$ and individual-level (random-effect) terms $\boldsymbol{\eta}_i$. In this case, the parameters to be inferred are
\begin{equation}
    \boldsymbol{\Theta} = \{ \boldsymbol{\theta}_{pop}, \boldsymbol{\Omega}, \boldsymbol{\Sigma} \},
\end{equation}
which includes the vector $\boldsymbol{\theta}_{pop}$ representing the population-level parameters, the matrix $\boldsymbol{\Omega}$ which expresses the between-patient inter-individual variability (IIV), and the residual error matrix $\boldsymbol{\Sigma}$.
Although Eq.~\ref{eq:llh} specifies the targeted likelihood, to include these random effects, common PK algorithms instead employ the joint likelihood 
\begin{equation}\label{eq:joint-llh}
    p(\mathbf{Y}, \boldsymbol{\eta} \mid \boldsymbol{\Theta}) = \prod_i p(\boldsymbol{y}_i, \boldsymbol{\eta}_i \mid \boldsymbol{\Theta}), 
\end{equation}
from which the original likelihood (Eq.~\ref{eq:llh}) is obtained by integrating over $\boldsymbol{\eta}$, where $\boldsymbol{\eta} = (\boldsymbol{\eta}_1,\dots,\boldsymbol{\eta}_M)$ describes the individual-level parameters.

FOCEI and SAEM are two commonly used frequentist methods for nonlinear mixed-effects models whose algorithms use this joint likelihood. 
Both methods use an alternating-update approach to estimate the parameters, the exact method by which varies between the two. 
On the other hand, MCMC and HMC, two commonly used Bayesian inference methods, use this likelihood to update all parameters jointly. We will discuss these different methods now.

\subsubsection{FOCEI}

FOCEI (see \cite{sheiner1980evaluation,wang2007derivation}) exploits the fact that the target likelihood (Eq.~\ref{eq:llh}) is simply the joint likelihood (Eq.~\ref{eq:joint-llh}), but with the random effect variables integrated out, that is,
\begin{equation}\label{eq:nlme-llh}
    L(\boldsymbol{\Theta}) =  \prod_i \int p(\boldsymbol{y}_i, \boldsymbol{\eta}_i \mid \boldsymbol{\Theta}) d\boldsymbol{\eta}_i = \prod_i \int p(\boldsymbol{y}_i \mid \boldsymbol{\eta}_i,\boldsymbol{\theta}_{pop},\boldsymbol{\Sigma}) p(\boldsymbol{\eta}_i\mid \boldsymbol{\Omega}) d\boldsymbol{\eta}_i.
\end{equation}
Note that the simplification in the final expression is due to the independence between parameters: first, between the output and the IIV matrix given the random effects, and second, as the random effects depend only on the IIV covariance itself. However, the nonlinearity of these mixed-effects models renders this integral analytically intractable. To handle this, the terms inside the integral are approximated such that the integral becomes tractable. The FOCEI algorithm does this in multiple steps. First, given the current estimate of the parameters $\boldsymbol{\hat{\Theta}}$, an estimate for the random effects terms $\hat{\boldsymbol{\eta}}_i$ is computed for each patient $i$. This is achieved by maximising the posterior distribution of the random effects
\begin{equation}\label{eq:focei-eta}
    \hat{\boldsymbol{\eta}}_i = \text{arg} \max_{\boldsymbol{\eta}_i} \enspace p(\boldsymbol{\eta}_i \mid \boldsymbol{y}_i, \boldsymbol{\hat{\Theta}}) = \text{arg} \max_{\boldsymbol{\eta}_i} \enspace p(\boldsymbol{y}_i \mid \boldsymbol{\eta}_i,\boldsymbol{\hat{\theta}}_{pop},\boldsymbol{\hat{\Sigma}}) p(\boldsymbol{\eta}_i \mid \boldsymbol{\hat{\Omega}}),
\end{equation}
where the second expression comes from the application of Bayes' rule (and where the normalising constant has been dropped). We also note that this expression aligns with that in Eq.~\ref{eq:nlme-llh}. Next, the expression is approximated by linearising about the current estimate $\boldsymbol{\hat{\eta}}_i$, making the expression Gaussian and therefore the integral tractable \cite{wang2007derivation}. This gives an analytical expression for the likelihood, which can then be used to update the parameter estimates $\boldsymbol{\hat{\Theta}}$ via standard optimisation techniques. The fact that in Eq.~\ref{eq:focei-eta}, the MAP of the posterior distribution $p(\boldsymbol{\eta}_i \mid \boldsymbol{y}_i, \boldsymbol{\hat{\Theta}})$ is used to estimate the random effects terms is why these are commonly referred to as empirical \textit{Bayes} estimates. 

One thing to note is that by targeting the MAP while estimating the random effects $\boldsymbol{\eta}$, we incorporate a prior on $\boldsymbol{\eta}$, effectively introducing regularisation into the optimisation problem (see Fig.~\ref{fig:comp_bayes_freq}). This phenomenon is, in fact, what is often referred to in pharmacometrics as \textit{shrinkage}, as it encourages each patient's random effect to shrink towards zero. In essence, shrinkage can be viewed as the regularisation of the random effects terms in mixed-effects models.

\subsubsection{SAEM}

SAEM targets the likelihood optimisation problem from a different direction. Instead of integrating out the random effects variable like in FOCEI, it instead uses a method called expectation-maximisation that explicitly treats these random effects as latent variables in order to maximise the likelihood with respect to the parameters \cite{delyon1999convergence,kuhn2005maximum,lavielle2014mixed}. The algorithm consists of two steps:

\begin{enumerate}
    \item[(1)] \textit{Expectation}: Given the current estimates for the parameters $\boldsymbol{\hat{\Theta}}$, the expected value of the joint likelihood with respect to the posterior of the random effects terms is computed, 
    \begin{equation}
    Q(\boldsymbol{\Theta},\boldsymbol{\hat{\Theta}}) := \mathbb{E}_{p(\boldsymbol{\eta}\mid \mathbf{Y},\boldsymbol{\hat{\Theta})}}[\log p(\mathbf{Y},\boldsymbol{\eta}\mid \boldsymbol{\Theta})] = \int \log p(\mathbf{Y},\boldsymbol{\eta}\mid \boldsymbol{\Theta}) p(\boldsymbol{\eta}\mid \mathbf{Y},\boldsymbol{\hat{\Theta}}) d\boldsymbol{\eta}.
\end{equation}
    The result is a function of both the parameters $\boldsymbol{\Theta}$ and the corresponding estimates $\boldsymbol{\hat{\Theta}}$, which is commonly denoted as $Q(\boldsymbol{\Theta},\boldsymbol{\hat{\Theta}})$. 
    This step effectively takes into account the range of possible values for $\boldsymbol{\eta}$, rather than focusing on a point estimate, as is done in FOCEI. Note that within this step, the nonlinearity of the model requires that numerical techniques, in this case MCMC, are employed to approximate the integral.
    \item[(2)] \textit{Maximisation}: here, the function $Q(\boldsymbol{\Theta},\boldsymbol{\hat{\Theta}})$ defined by the expectation step is maximised to give a new estimate for the global parameters $\boldsymbol{\hat{\Theta}}$. 
    Based on the underlying theory for EM, maximising this function corresponds to an improvement in the likelihood function, with the algorithm stopping once convergence has been reached. Unlike exact EM, SAEM does not guarantee an increase at every iteration, but it converges under mild conditions.
\end{enumerate}

Since the algorithm itself is more involved, it can take longer to run SAEM compared to FOCEI \cite{sukarnjanaset2018evaluation}. However, if the model with respect to $\boldsymbol{\eta}$ is highly nonlinear, then SAEM can be more accurate, as it does not approximate the model with a normal distribution, as is done in FOCEI. 

\subsubsection{MCMC/HMC}

Bayesian-based methods, such as MCMC/HMC, take a different approach altogether. Instead of iterating between updating steps during computation, all parameters are estimated jointly through the target posterior distribution. This can be expressed for mixed-effects models as
\begin{equation}
    p(\boldsymbol{\Theta}\mid\mathbf{Y}) = \frac{p(\mathbf{Y}\mid\boldsymbol{\theta}_{pop},\boldsymbol{\Omega},\boldsymbol{\Sigma})p(\boldsymbol{\theta}_{pop})p(\boldsymbol{\Omega})p(\boldsymbol{\Sigma})}{p(\mathbf{Y})}.
\end{equation}
However, this posterior distribution may be difficult to compute in practice, oftentimes requiring approximation methods, as is the case for nonlinear mixed-effects models.

MCMC is a method to approximate the posterior distribution using samples. This method is based on Markov chains, a mathematical process that describes the way the state of a system changes, where the probability of transitioning from one state to the next depends only on the current state. If the transition between states is designed in a specific way, then after sufficiently many iterations, the visited states are representative of samples from the posterior, and their empirical distribution provides an approximation to the posterior distribution \cite{metropolis1953equation,hastings1970monte}.

Despite its popularity and wide applicability, there are many problems with standard MCMC and its variants in practice. In particular, oftentimes they exhibit random-walk behaviour, which reduces the quality of the sampling, especially in high-dimensional problems \cite{roberts1997weak}. 
As the aim is to generate samples that are representative of the distribution of interest, if the transition is not well designed, the samples may not fully capture the distribution, potentially biasing the resulting inferences.

HMC is a form of MCMC where the proposal scheme uses information about the posterior distribution to improve sampling. Instead of generating a proposal at random based on the current sample, HMC instead uses the simulation of Hamiltonian trajectories to propose a new state, in a way akin to a particle moving around an energy well \cite{duane1987hybrid,neal2011mcmc,betancourt2017conceptual}. This approach results in better transition proposals, therefore helping produce samples that better represent the posterior distribution. An extension of HMC, based on an adaptive algorithm known as NUTS \cite{hoffman2014no} as implemented in Stan, is used for the Bayesian analysis in this paper.

\section{Methods}

Having discussed the theoretical differences and algorithms behind typical frequentist and Bayesian methods for nonlinear mixed-effects models, we now turn to understanding how these differences are reflected in practice. To do this, we consider a classic problem in pharmacometrics: inferring the parameters that govern a drug's absorption, distribution, and elimination from measured concentration-time data. 

\subsection{Compartmental PK models}\label{sec:model}

\begin{figure}
    \centering
    \begin{tikzpicture}[node distance=2cm, auto]

    % Compartment 1 (Central Compartment)
    \node[draw, rectangle, rounded corners, text width=2.5cm, minimum width=2.5cm, minimum height=2cm, align=center, fill=teal!10] (c1) {Central \\Compartment ($V_C$)};
    
    % Compartment 2 (Peripheral Compartment 1)
    \node[draw, rectangle, rounded corners, text width=2.5cm, minimum width=2.5cm, minimum height=2cm, right=1.3cm of c1, align=center, fill=teal!10] (c2) {Peripheral \\Compartment 1 \\($V_{P_1}$)};

    % Compartment 3 (Peripheral Compartment)
    \node[draw, rectangle, rounded corners, text width=2.5cm, minimum width=2.5cm, minimum height=2cm, left=1.3cm of c1, align=center, fill=teal!10] (c3) {Peripheral \\Compartment 2 \\($V_{P_2}$)};

    % Dotted box around Central and Peripheral Compartment 1
    \node[draw, densely dashed, rounded corners, inner xsep=0.0cm, inner ysep=0.8cm, fit=(c1) (c2), name=twocomp] {};

    \node[anchor=north east,font=\small] at (twocomp.north east) {Two-compartment};

    % Dotted box around all compartments
    \node[draw, dashed, rounded corners, inner xsep=0.3cm, inner ysep=1.2cm, fit=(c1) (c2) (c3), name=threecomp] {};

    \node[anchor=north east,font=\small] at ($(threecomp.north west)+(+3.3cm,0)$) {Three-compartment};
    
    % Elimination from Compartment 1
    \draw[-Latex] (c1.south) -- ++(0,-0.6) node[midway, left] {$C_L$};
    
    % Inter-compartmental transfer 
    \draw[Latex-Latex] ($(c1.east)+(0,0em)$) -- ($(c2.west)+(0,0em)$) node[midway, above] {$Q_1$};

    % Inter-compartmental transfer 
    \draw[Latex-Latex] ($(c1.west)+(0,0em)$) -- ($(c3.east)+(0,0em)$) node[midway, above] {$Q_2$};
    
    % Injection into Compartment 1 (D)
    \draw[-Latex] (c1.north) ++(0,0.6) -- (c1.north) node[midway, right] {$R_{in}(t)$};
    
    \end{tikzpicture}
    \caption{Illustration of two- and three-compartment pharmacokinetic models. Both models include the input dosage rate $R_{in}(t)$ that is an input into the central compartment with volume $V_C$ and clearance $C_L$. The two-compartment model includes one peripheral compartment with volume $V_{P_1}$ and intercompartmental clearance $Q_1$. The three-compartment model also includes a second peripheral compartment parameterised by $V_{P_2}$ and $Q_2$.}
    \label{fig:comp-pk}
\end{figure}
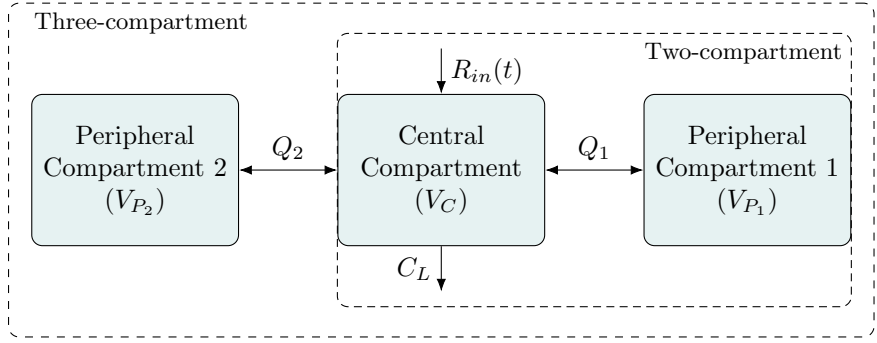

We consider both two- and three-compartment PK models in which the drug is administered intravenously into the central compartment. Fig. \ref{fig:comp-pk} gives a visual representation of this process. 
The drug enters the central compartment, the volume of which is denoted $V_C$, at a possibly time-varying input rate $R_{in}(t)$.  
The drug then moves between either one or two peripheral compartments with volumes $V_{P_1}$ and $V_{P_2}$.
The exchange between the central compartment and the peripheral compartments is characterised by the intercompartmental clearance $Q_1$ for the first compartment and $Q_2$ for the second compartment, if present in the model. Lastly, excretion of the drug from the system via the central compartment is modelled by the clearance rate parameter $C_L$. 

Both the two- and three-compartment PK models can be described by a set of ordinary differential equations (ODEs) that model the amount of drug in each compartment. These equations can be simplified via a reparameterisation (see Appendix A) that simplifies inference. Under this reparameterisation, the two-compartment model is defined by
\begin{subequations}\label{eq:ode-model-2cmpt}
    \begin{align}
        \frac{dA_{c}(t)}{dt} &= R_{in}(t) - ( k_{10} + k_{12}) \cdot A_{c}(t) + k_{21} A_{1}(t),    \\
        \frac{dA_{1}(t)}{dt} &=  k_{12} A_{c}(t) - k_{21} A_{1}(t),
    \end{align}
\end{subequations}
and the three-compartment model by,
\begin{subequations}\label{eq:ode-model-3cmpt}
    \begin{align}
        \frac{dA_{c}(t)}{dt} &= R_{in}(t) - ( k_{10} + k_{12} + k_{13}) \cdot A_{c}(t) + k_{21} A_{1}(t)  + k_{31} A_{2}(t),    \\
        \frac{dA_{1}(t)}{dt} &=  k_{12} A_{c}(t) - k_{21} A_{1}(t),  \\
        \frac{dA_{2}(t)}{dt} &=  k_{13} A_{c}(t) - k_{31} A_{2}(t),  
    \end{align}
\end{subequations}
with the initial conditions $A_{c}(0) = A_{c}^{init}$, $A_{1}(0) = A_{1}^{init}$ and $A_{2}(0) = A_{2}^{init}$ indicating the amount of drug in the central, first and second peripheral compartments respectively. 

We model the system using a mixed-effects model. We describe the situation for the three-compartment model here, which is straightforward to reduce to the two-compartment case. For this model, the population-level (fixed-effect) parameters are 
\begin{equation}\label{eq:pop-params}
    \boldsymbol{\theta}_{pop} = (\theta_{C_L}, \theta_{V_C}, \theta_{Q_1}, \theta_{Q_2}, \theta_{V_{P_1}},\theta_{V_{P_2}})
\end{equation}
with associated individual-level (random effect) parameters for each individual $i=1,\dots,M$, 
\begin{equation}\label{eq:ind-params}
    \boldsymbol{\eta}_i = (\eta_{C_L,i},\eta_{V_C,i},\eta_{Q_1,i},\eta_{Q_2,i},\eta_{V_{P_1},i},\eta_{V_{P_2},i}).
\end{equation}
Furthermore, we assume independent random effects, so the IIV covariance matrix is diagonal,
\begin{equation}
    \boldsymbol{\Omega} = \operatorname{diag}\!\left(
        \omega_{C_L}^2,\, \omega_{V_C}^2,\, \omega_{Q_1}^2,\,
        \omega_{Q_2}^2,\, \omega_{V_{P_1}}^2,\, \omega_{V_{P_2}}^2 \right),
\end{equation}
and that the residual error combines additive and proportional terms \cite{mould2013basic}, giving 
$\boldsymbol{\Sigma} = \operatorname{diag}(\sigma_1^2,\, \sigma_2^2)$.
For the mixed-effects model, each individual has a set of patient-specific pharmacokinetic parameters
\begin{equation}
    \boldsymbol{\theta}_i = ( C_{L,i}, V_{C,i}, Q_{1,i}, Q_{2,i}, V_{P_1,i}, V_{P_2,i} )
\end{equation}
which relate to both population-level $\boldsymbol{\theta}_{pop}$ and individual-level $\boldsymbol{\eta}_i$ parameters. Take, for example, the systemic clearance for patient $i$ denoted $C_{L,i}$. This is related to both the population parameter $\theta_{C_L}$ (Eq.~\ref{eq:pop-params}) and the individual parameter for the patient $\eta_{C_L,i}$ (Eq.~\ref{eq:ind-params}) by
\begin{equation}
    C_{L,i} = \theta_{C_L}\cdot\exp\left( \eta_{C_L,i}\right).
\end{equation}
This same relationship applies to the other patient-specific pharmacokinetic parameters.
 
For both systems, the primary quantity of interest is the concentration of the drug in the central compartment for each patient $i$, defined as,
\begin{equation}\label{eq:meas-model}
    y_{i,t} = \frac{A_{c,i}(t)}{V_{C,i}}.
\end{equation}
We assume that the observations of the system are normally distributed, such that the relationship between the observations $\tilde{y}_{i,t}$ and the model output is described as
\begin{equation}\label{eq:obs-model}
    \tilde{y}_{i,t} \sim \mathcal{N}(y_{i,t}(\boldsymbol{\theta}_i), \sigma_1+y_{i,t}\cdot\sigma_2)
\end{equation}
where $y_{i,t}(\boldsymbol{\theta}_i)$ denotes the output of the ODE system for patient $i$ at observation time-point $t$ given the patient-specific parameters $\boldsymbol{\theta}_i$. Note that here and for the remainder of the paper, the normal distribution is specified with standard deviation, as opposed to variance, to be consistent with the Bayesian inference software used in the results.

\subsection{Experimental setup}

We evaluate the inference methods in two steps: first on simulated data, where known ground truth allows performance to be measured directly, and then on a real dataset, to assess the methods under realistic complexity. For all analysis we use open-source software and publicly available data to ensure access and reproducibility. We use the \texttt{R} programming language for all simulations \citep{r-core-team}. For frequentist methods, we use the open-source package \texttt{nlmixr2}. The probabilistic programming language \texttt{RStan} \cite{rstan} is used to conduct the Bayesian inference simulations. 

\subsubsection{Simulated data}
\begin{figure}[htb]
    \centering
    \includegraphics[width=1\linewidth]{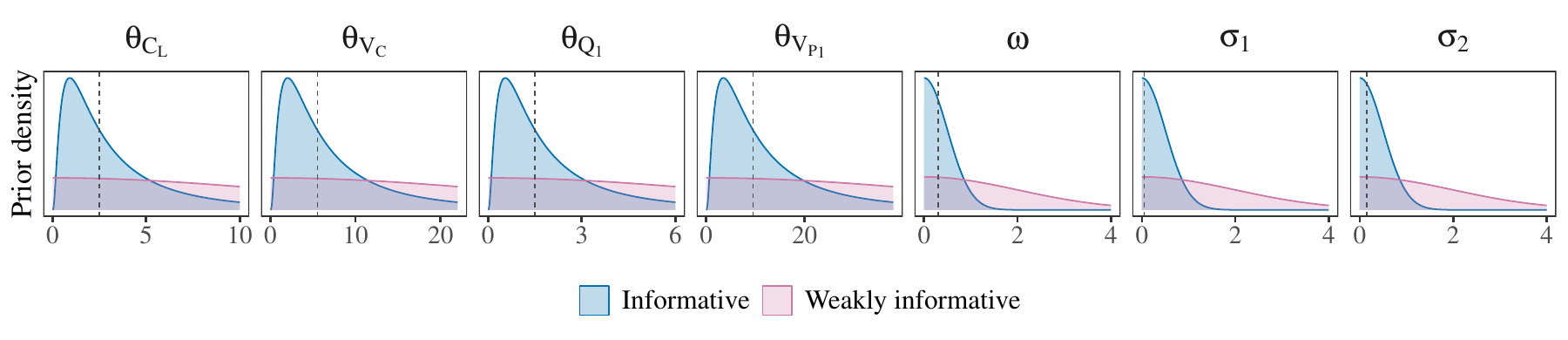}
    \caption{The two prior settings (informative and weakly informative) for each parameter to be inferred. Dashed vertical lines mark the true values}
    \label{fig:simdata_prior}
\end{figure}

For the simulated data experiments, we consider the two-compartment PK model with a single IV bolus input, and compare four inference methods: the frequentist algorithms FOCEI and SAEM, and the Bayesian sampler HMC/NUTS under a weakly informative and an informative prior. The two priors are illustrated in Fig.~\ref{fig:simdata_prior}, and comparing them isolates the influence of prior information. To emulate the input, we initialise the system at $A_{c,i}^{init}=100$mg, $A_{1,i}^{init}=A_{2,i}^{init}=0$mg and set $R_{in}(t)=0$.
Since the model is linear (Eq.~\ref{eq:ode-model-2cmpt}), its solution is available in closed form. For a single IV bolus, this is a biexponential, so no numerical ODE solver is needed, and the reported differences reflect the inference engine rather than the solver.

\noindent We simulate data from virtual patients whose parameters are drawn from a log-normal distribution about the typical population values
$$
(\theta_{C_L}, \theta_{V_C}, \theta_{Q_1}, \theta_{V_{P_1}})
= (2.5~\text{L/min},\, 5.5~\text{L},\, 1.5~\text{L/min},\, 9.5~\text{L}).
$$
The random effect corresponding to each parameter is drawn from $\mathcal{N}(0, \omega)$, where $\omega$ is the between-patient standard deviation on the log scale (baseline $\omega = 0.3$). 
Observations are generated from each simulated patient's concentration under the error model (Eq.~\ref{eq:obs-model}), with a fixed additive term $\sigma_1 = 0.05$ and a proportional term $\sigma_2$ (baseline $0.15$).

For each of the four inference methods, we estimate the population parameters $\boldsymbol{\theta}_{pop} = (\theta_{C_L}, \theta_{V_C}, \theta_{Q_1}, \theta_{V_{P_1}})$, the between-patient variability $\omega_k$ for each parameter $k$, and the two error parameters $\sigma_1$ and $\sigma_2$. 
We do this under a range of conditions defined around a baseline of $M = 30$, $N = 10$, $\sigma_2 = 0.15$, and $\omega = 0.3$. 
From this baseline, we vary one factor at a time: 
the sample size $M \in \{10, 30, 100\}$, 
the number of observations per patient $N \in \{5, 7, 10\}$, 
the proportional noise term $\sigma_2 \in \{0.05, 0.15, 0.30\}$, 
and the IIV standard deviation $\omega \in \{0.1, 0.3, 0.6\}$.

For each combination of factors, we simulate $30$ patient cohorts. 
For each simulated cohort and inference method, a single fit is performed. 
For FOCEI and SAEM, $\boldsymbol{\theta}_{pop}$ and $\sigma_1 = 0.05$ are initialised at their true values, and $\omega$ and $\sigma_2$ at their baseline values ($\omega = 0.3$, $\sigma_2 = 0.15$). 
For the Bayesian fits, the default prior-based initialisation is used, and four parallel chains are run for $2000$ iterations each, discarding the first $1000$ as warmup, with non-converged HMC fits re-seeded up to six times.

\subsubsection{Real-world data}
For the real-world experiments, we consider a publicly available remifentanil dataset \citep{minto1997influence}. The data come from a longitudinal study on the concentration of the remifentanil drug in $M=65$ healthy adults during and after intravenous infusion. The length of drug administration, the number of observations, the sampling frequency, and the maximum observed concentration vary based on the individual. 
The dosage input rate $R_{in}(t)$ \textmu g/min is modelled as a time-dependent input step function,
\begin{equation}
    R_{in}(t) = \begin{cases}
        R_d & \text{if } t \leq T_d \\
        0 & \text{if } t > T_d
    \end{cases}
\end{equation}
where $R_d$ \textmu g/min is the dosage rate. The drug concentration in the bloodstream is measured at several time points from the dosing start until the drug has been eliminated from the system. The initial amount of drug in each compartment is zero. 

For both two- and three-compartment models, we compare the frequentist SAEM method to the Bayesian HMC/NUTS sampler with the following weakly informative priors
\begin{align*}\label{eq:priors-uninform}
    \boldsymbol{\theta}_{pop} &\sim \text{logNormal}(\log(5),1), \\ \boldsymbol{\Omega} &\sim \text{logNormal}(\log(0.05),0.25), \\ \boldsymbol{\Sigma} &\sim \text{Half-Normal}(0,1).
\end{align*}
For the frequentist analysis, the fit is repeated 100 times, with the final estimates chosen based on the best Akaike information criterion (AIC) value. For the Bayesian analysis, eight parallel chains were run for 5000 iterations each, removing the first 1000 iterations as warmup. 

\noindent Additionally, we investigate the effect of using more informative priors,
\begin{align*}
    \theta_{C_L}      &\sim \text{logNormal}(\log(3),0.2),
    &\qquad \theta_{V_C}   &\sim \text{logNormal}(\log(5),0.2), \\ 
    \theta_{Q_1}      &\sim \text{logNormal}(\log(0.05),0.2),
    &\qquad \theta_{Q_2}   &\sim \text{logNormal}(\log(2),0.2), \\
    \theta_{V_{P_1}}  &\sim \text{logNormal}(\log(5),0.2),
    &\qquad \theta_{V_{P_2}} &\sim \text{logNormal}(\log(10),0.2), \\
    \boldsymbol{\Sigma} &\sim \text{Half-Normal}(0,0.2),
    &\qquad \boldsymbol{\Omega} &\sim \text{logNormal}(\log(0.05),0.1),
\end{align*}
on the posterior distribution of the three-compartment model.

\section{Results}\label{sec:results}

\subsection{Comparison of methods on simulated data}

\begin{figure}[htb]
    \centering
    \includegraphics[width=1\linewidth]{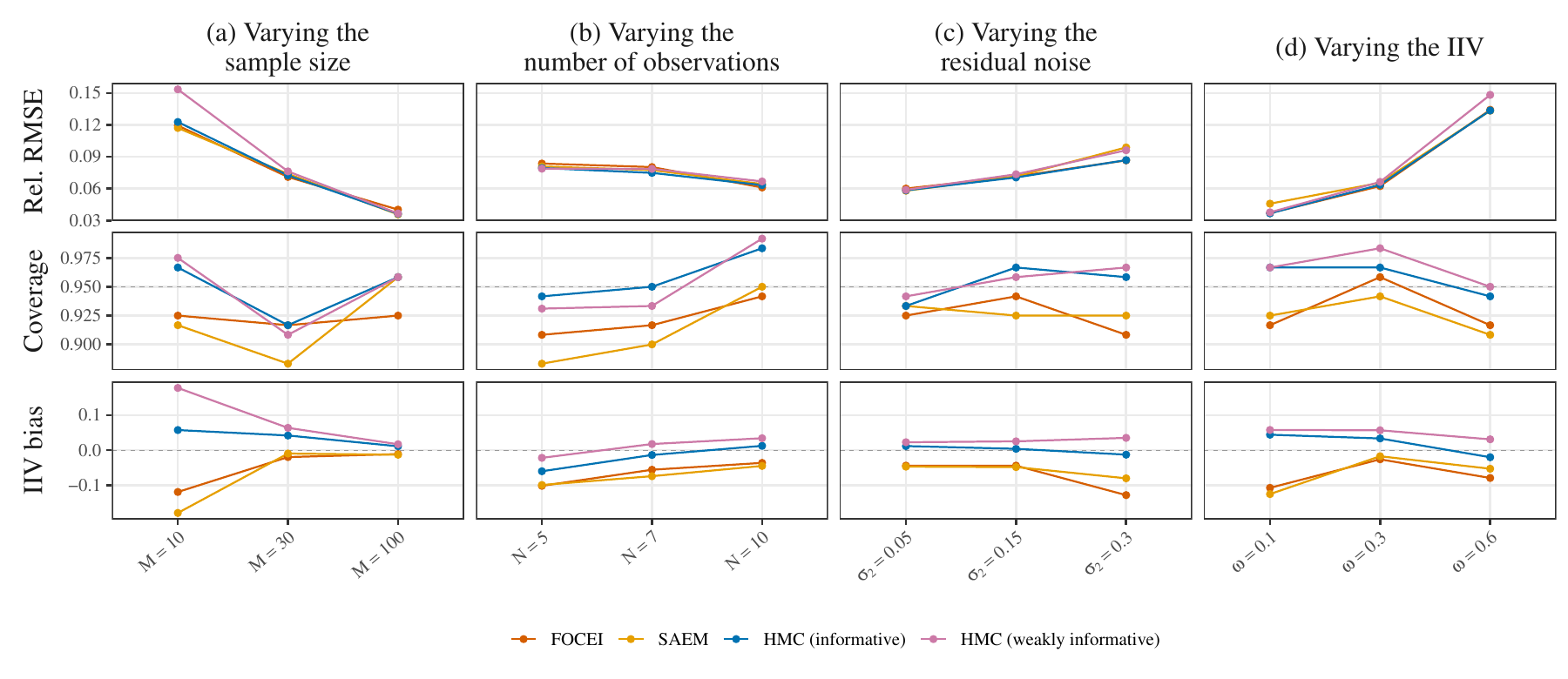}
    \caption{Performance of the four inference methods (FOCEI, SAEM, informative prior HMC, weakly informative prior HMC). Top row: root mean square error (RMSE, normalised by the true value) of the population estimates $\boldsymbol{\theta}_{pop}$.
    Middle row: coverage, i.e., the fraction of inference runs whose $95\%$ confidence or credible intervals contained the true value.
    Bottom row: relative bias in the inter-individual variability (IIV) $\omega$, the standard deviation of the random effects that make patients differ from the mean population value. 
    A value $X$ means the inferred $\omega$ is over- ($X>0$) or under-estimated ($X<0$) by a fraction $|X|$ of the truth. 
    The points are averaged over the four model parameters and 30 simulated patient cohorts, using only converged fits
    }
    \label{fig:simdata_sum}
\end{figure}

Fig.~\ref{fig:simdata_sum} summarises the performance of the considered inference methods across all varying factors.
For point estimation (top row of Fig.~\ref{fig:simdata_sum}), the methods perform almost identically as their relative root mean square error (RMSE) curves nearly coincide in every panel. 
In our results, differences in RMSE are thus not driven by the inference method, but rather by the varying data factors, notably the sample size $M$ and the IIV $\omega$. 
The middle row of Fig.~\ref{fig:simdata_sum} shows coverage, defined here as the fraction of runs whose $95\%$ interval contains the true value. 
FOCEI and SAEM tend to under-cover, with their Wald confidence intervals lying below the nominal $95\%$, meaning they mildly underestimate the uncertainty in their estimates. The coverage of HMC credible intervals is higher than that of FOCEI and SAEM. 
This difference is, for example, noticeable at small sample sizes and high residual noise. 
The bottom row of Fig.~\ref{fig:simdata_sum} shows that the methods also separate in their inference of the between-patient variability $\omega$.
FOCEI and SAEM lie below the zero-bias line in every panel, meaning they consistently underestimate $\omega$. 
The HMC methods, which lie above FOCEI and SAEM throughout, instead tend to overestimate $\omega$. Of these, the informative prior lies closer to the zero-bias line than the weakly informative prior.

\begin{figure}[htb]
    \centering
    \includegraphics[width=1\linewidth]{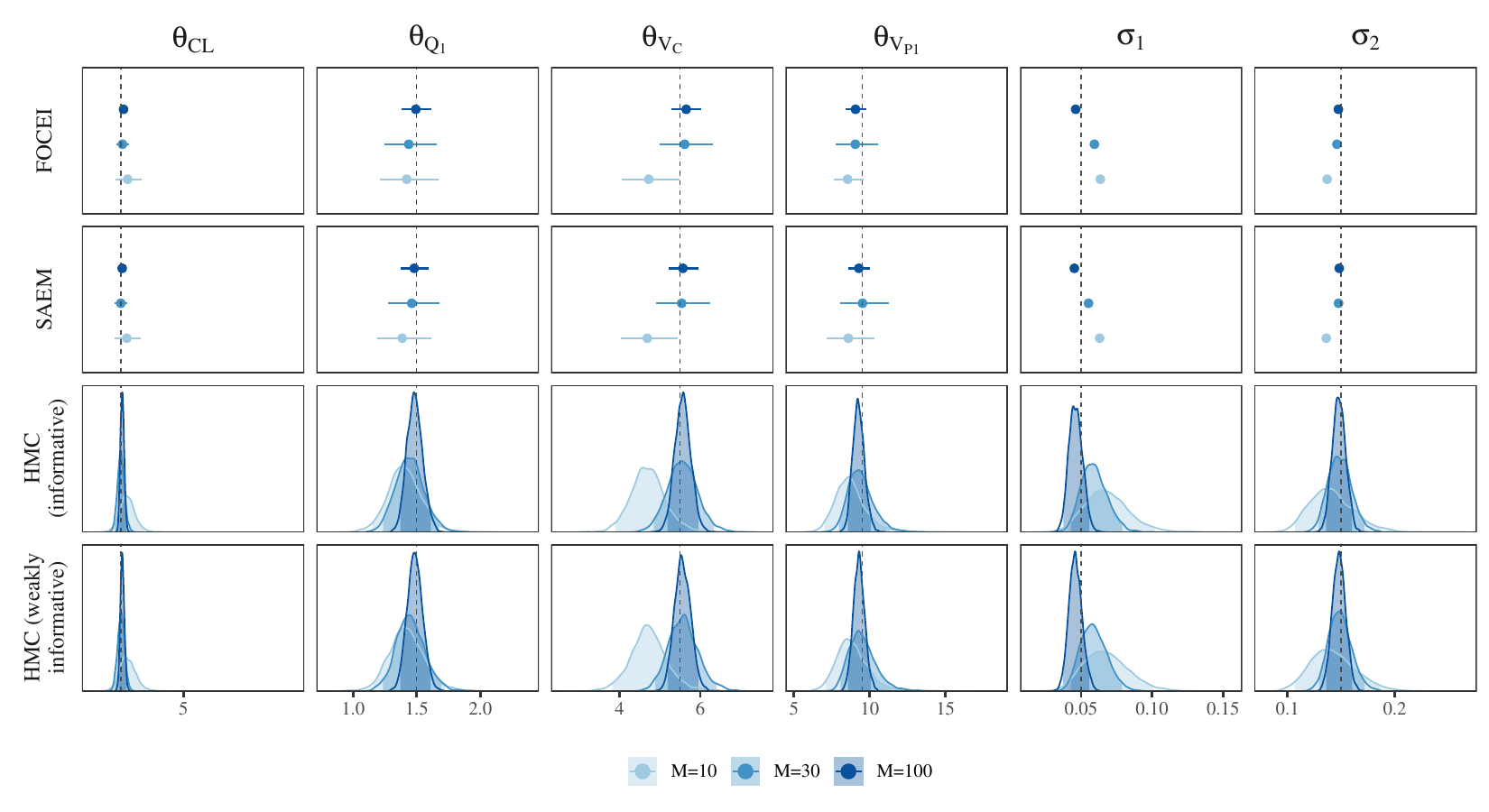}
    \caption{Inferred population and error parameters for a single fit as the sample size $M$ varies. 
    The other factors are held at baseline ($N=10$, $\sigma_2=0.15$, $\omega=0.3$).
    The frequentist methods (FOCEI, SAEM) are shown as point estimates with 95\% Wald confidence intervals; no interval is shown for the residual-error terms $\sigma_1,\sigma_2$, for which nlmixr2 does not report standard errors. 
    The Bayesian methods (HMC with informative and weakly informative priors) are shown as posterior densities, with the central 95\% of each shaded. 
    Dashed vertical lines mark the true values; colour encodes the sample size
    }
    \label{fig:simdata_varyM}
\end{figure}

\begin{figure}[H]
    \centering
    \includegraphics[width=1\linewidth]{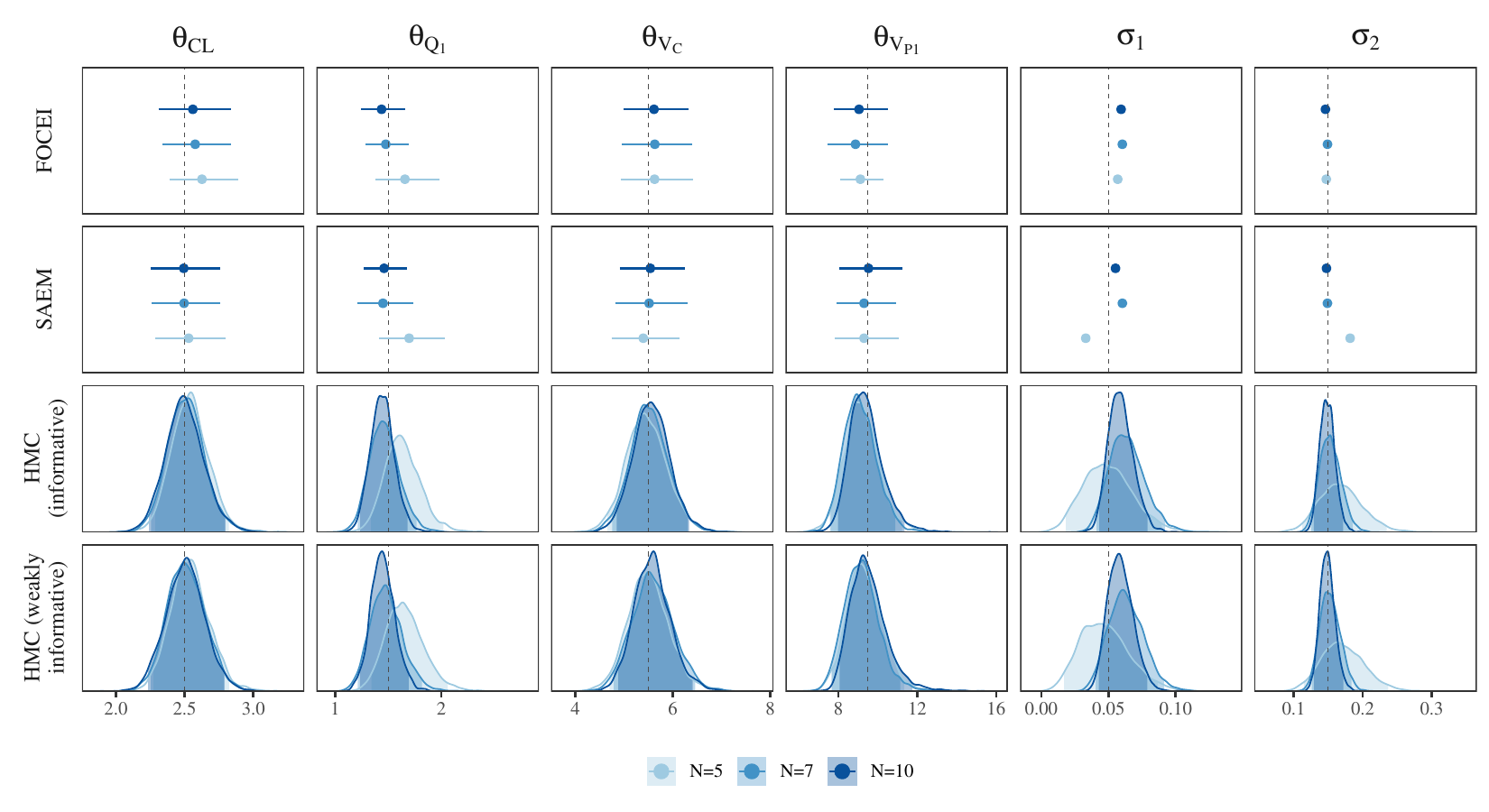}
    \caption{Inferred population and residual-error parameters for a single fit as the number of observations per patient $N$ varies. 
    The other factors are held at baseline ($M=30$, $\sigma_2=0.15$, $\omega=0.3$).
    The frequentist methods (FOCEI, SAEM) are shown as point estimates with 95\% Wald confidence intervals; no interval is shown for the residual-error terms $\sigma_1,\sigma_2$, for which nlmixr2 does not report standard errors. 
    The Bayesian methods (HMC with informative and weakly informative priors) are shown as posterior densities, with the central 95\% of each shaded. 
    Dashed vertical lines mark the true values; colour encodes the number of observations per patient $N$
    }
    \label{fig:simdata_varyN}
\end{figure}

Figs.~\ref{fig:simdata_varyM} and~\ref{fig:simdata_varyN} show these trends for a single fit (of $M$ simulated patients) as the sample size $M$ and the number of observations per patient $N$ vary, respectively. 
Estimates from each inference method concentrate around the truth as data increase, and the Wald intervals and central-$95\%$ posteriors largely agree once data are rich.
The two Bayesian posteriors are nearly indistinguishable, indicating that the data dominate both priors.

The four methods also differ sharply in computational cost. 
The median runtime per fit is $\sim$3--4~s for FOCEI and SAEM, versus $\sim$18~s for the Bayesian methods, which thus incur roughly a five- to six-fold penalty for full posterior sampling.
All figures report converged HMC fits only. Of the $360$ fits per prior, $6$ (informative) and $9$ (weakly informative) required re-seeding to reach convergence (no divergent transitions, $\widehat{R} < 1.05$ \cite{vehtari2021rank}); one weakly informative fit failed to converge within six re-seeds and was excluded. The frequentist fits required no re-runs.

\subsection{Case study: Remifentanil}

\begin{figure}[htbp]
    \centering
        \includegraphics[width=\linewidth]{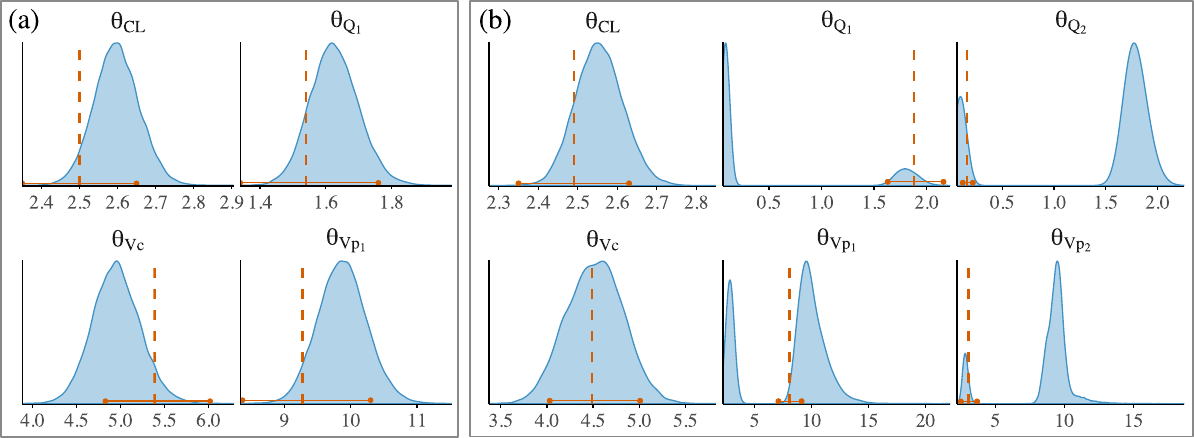}
    \caption{Posterior densities for population parameters for the (a) two- and (b) three-compartment models. Frequentist (SAEM) results are displayed in orange, with the parameter estimate shown by the dashed line and the 95\% confidence interval by the horizontal line}\label{fig:posteriors-two-three}
\end{figure}

Fig.~\ref{fig:posteriors-two-three} shows the frequentist estimates (including 95\% confidence intervals) and Bayesian posteriors for the population parameters for the two- and three-compartment models. The resulting parameter estimates for the different methods cover a similar interval, with the frequentist method focusing on one mode rather than both in the three-compartment case. While the central compartment parameters ($\theta_{C_L}$ and $\theta_{V_C}$) remain relatively consistent between the models, the additional compartment introduces multimodality for both the intercompartmental clearance and peripheral volume, showing that two regions of the posterior are supported by the data. This arises from the structural nonidentifiability induced by the three-compartment model: the clearance parameters are interchangeable, such that $\theta_{Q_1}$ or $\theta_{Q_2}$ may correspond to the smaller clearance rate, with a corresponding interchangeability between the peripheral volumes. Importantly, we note that this nonidentifiability can be difficult to capture when taking a frequentist approach \cite{lavielle2016we,cassidy2026nonparametric}.

\begin{figure}[h!]
    \centering
    \includegraphics[width=\linewidth]{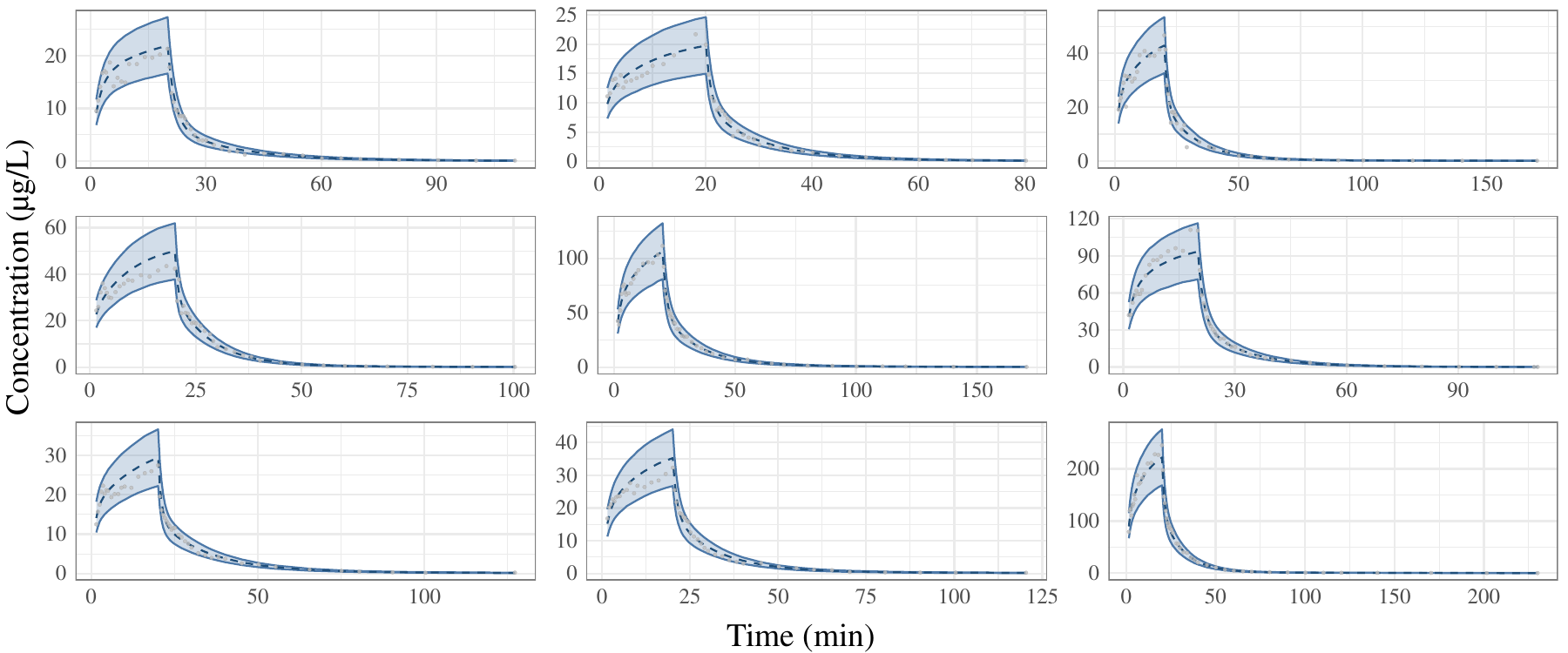}
    \includegraphics[width=\linewidth]{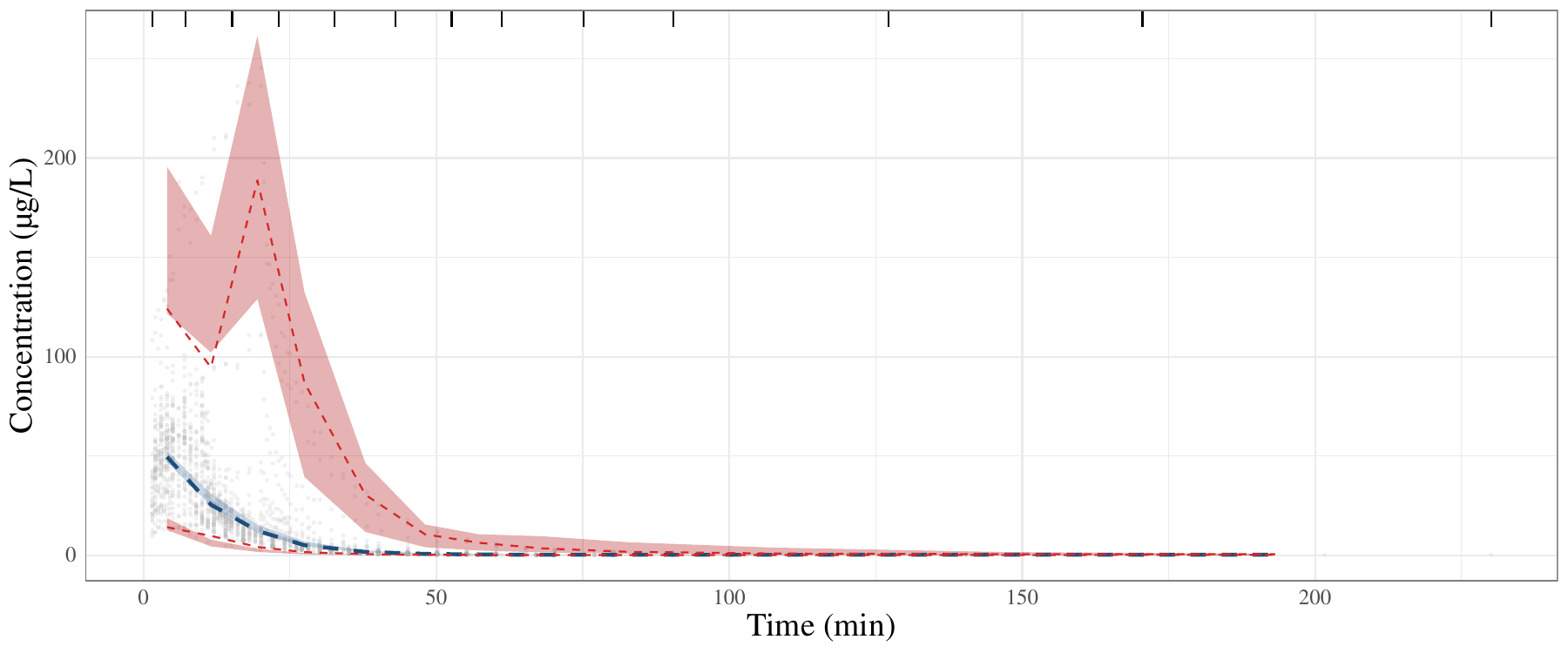}
    \caption{
    Visual diagnostics commonly used to assess agreement between model predictions and observed time-series data. Bayesian methods use posterior predictive checks (PPC), which compare forward predictions simulated from the posterior against the observed data (upper plots, nine randomly selected individuals). Frequentist methods instead often use visual predictive checks (VPC) across the whole population, comparing the median and selected upper and lower quantiles of the simulated predictions against all observations (lower plot). Throughout, the median and the 95\% prediction interval are shown in blue and the observations in grey. In the VPC, the 5\% and 95\% quantiles and their corresponding 95\% prediction intervals are additionally shown in red.
    }
    \label{fig:vpc-ppc-comp}
\end{figure}

Fig.~\ref{fig:vpc-ppc-comp} visualises the predictive distributions representative of the different approaches for the three-compartment model. There is close agreement between the predictive distributions and the observations, showing that each method adequately captures the individual dynamics. The posterior predictive checks (PPC) for Bayesian analysis typically focus on each individual's predictions, here showing the 95\% credible interval for the predictive median, where different maximum concentration levels are easily captured. The frequentist visual predictive check (VPC) shows the 95\% confidence interval around the 5\%, 95\% and median for the population, where the difference in concentration peaks between individuals is instead reflected through the predicted quantile estimates.

Regardless of the consistency in predicted dynamics, the multimodality introduced by the additional compartment shows up in the convergence diagnostics for the Bayesian methods. Since the simulated chains have converged to the different areas of the posterior, the reported $\widehat{R}$-statistic is larger than the standard threshold of 1.05 \cite{vehtari2021rank}, resulting in diagnostic warnings. This is ultimately a consequence of the symmetry of the problem, but it is important to resolve in practice. We therefore seek to include additional information in the inference problem to search for local practical identifiability. One approach would be to constrain the parameters to enforce this mathematically (e.g., setting $k_{21} > k_{31}$). Instead, we take a Bayesian approach to addressing this by introducing informative priors. 

\begin{figure}[h]
    \centering
    \includegraphics[width=\linewidth]{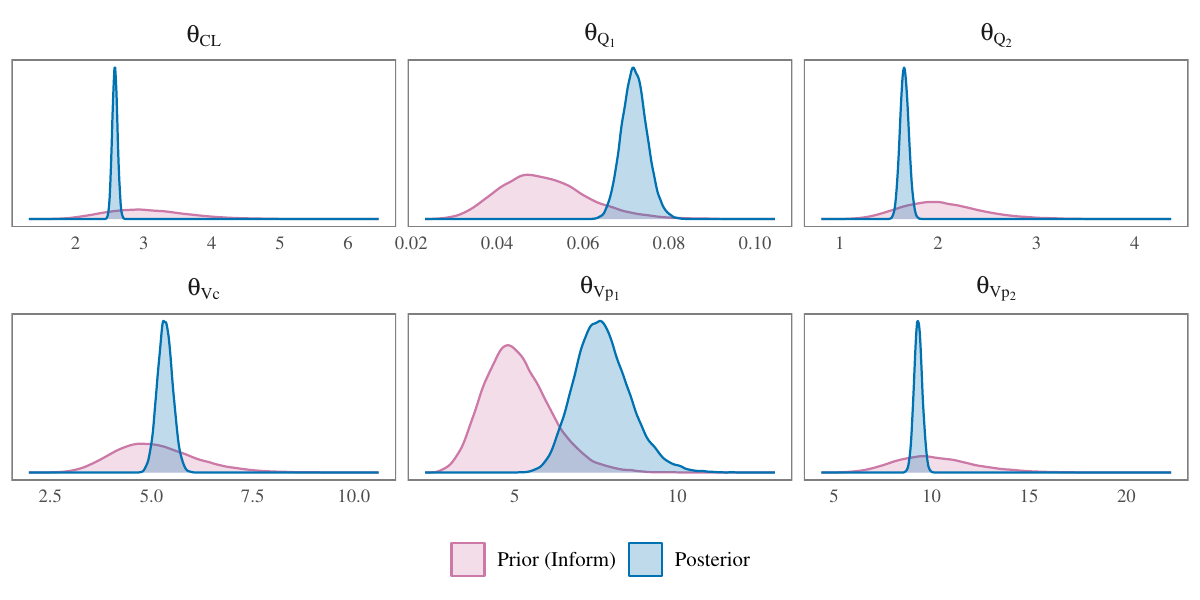}
    \caption{Comparison of prior (pink) and posterior (blue) distributions for the three-compartment model when using more informative priors that centre the solution at one mode}
    \label{fig:three-compt-full-patient-inform}
\end{figure}

Fig.~\ref{fig:three-compt-full-patient-inform} shows the posterior distributions for the three-compartment system when using informative priors. The prior is also shown for each parameter. We can see that the multimodality has now been removed from the posterior, such that we are effectively targeting a specific region based on our prior beliefs. Here, we show how knowledge of the system can be incorporated into the prior distributions to assist with the sampling, a technique that is an advantage of using a Bayesian approach.

\section{Conclusion}

Both Bayesian and frequentist approaches are popular methods for inference in pharmacometrics. 
In this work, we performed a side-by-side comparison of the two approaches, considering both their theoretical underpinnings and practical implementation, and highlighting their key similarities and differences. 
Analysis of simulated data enabled direct comparison of the methods against known parameter values and demonstrated the performance across a range of practically motivated scenarios, including changes in sample size, observation sparsity, residual proportional error, and the magnitude of between-patient variability. 
The real-world analysis complemented this by illustrating how each approach performs in practice for both the two- and three-compartment models, including how inference and uncertainty are typically presented for the parameter estimates and model predictions. We demonstrated that the additional compartment in the three-compartment model introduced nonidentifiability in the model, and how Bayesian methods were able to characterise this through the full posterior, offering information that was difficult to capture using a standard frequentist approach alone.
Further to this, we highlighted some of the practical advantages of each approach: frequentist methods generally offer faster simulation and computational times, whereas Bayesian methods allow prior distributions to be incorporated in a coherent, probabilistic way, and enable direct visualisation of the full posterior distribution. 
Overall, this work provides a practical point of reference for researchers considering the relative advantages, limitations, and implementation of Bayesian and frequentist approaches, and may serve as a guide for researchers interested in transitioning between them.

\backmatter

% \bmhead{Supplementary information}

% \bmhead{Acknowledgements}

\section*{Declarations}

\bmhead{Funding}
The work was partially supported by the Kjell och M{\"a}rta Beijer Foundation, which supported SH and JAC. 

\bmhead{Competing interests} 
The authors have no competing interests to declare that are relevant to the content of this article.

\bmhead{Ethics approval and consent to participate}
Not applicable.

\bmhead{Consent for publication}
Not applicable.

\bmhead{Data availability}

The data used for the simulation and real-world experiments can be found on \texttt{GitHub}: \url{https://github.com/jodie-c/freq-bayes-poppk}. The remifentanil dataset was originally published in \cite{minto1997influence} and can also be downloaded in the original format from \url{https://monolixsuite.slp-software.com/monolix/2024R1/remifentanil-data-set}

\bmhead{Code availability}
All code is available on \texttt{GitHub}: \url{https://github.com/jodie-c/freq-bayes-poppk}.

\bmhead{Author contributions}

Conceptualization: JAC, SH; Methodology: JAC, HW, LY, SH; Formal analysis and investigation: JAC, HW, LY, SH; Visualisation: JAC, SH. Writing - original draft preparation: JAC, SH; Writing - critical review and editing: JAC, HW, LY, TC, SH; Funding acquisition: JAC, SH.

\newpage

\begin{appendices}

\section{Appendices}

\subsection{Appendix A: PK Compartment Reparameterisation}\label{sec:app-reparam}

Both the two- and three-compartment PK models are described by a set of ODEs. This is expressed mathematically for the two-compartment system as
\begin{subequations}\label{eq:ode-model-2cmpt-og}
    \begin{align}
        \frac{dA_c(t)}{dt} &= R_{in}(t) - \left( \frac{C_L}{V_C} + \frac{Q_1}{V_C}\right) \cdot A_c(t) + \frac{Q_1}{V_{P_1}} A_1(t),    \\
        \frac{dA_1(t)}{dt} &=  \frac{Q_1}{V_C} A_c(t) - \frac{Q_1}{V_{P_1}} A_1(t),
    \end{align}
\end{subequations}
and for the three-compartment system as 
\begin{subequations}\label{eq:ode-model-3cmpt-og}
    \begin{align}
        \frac{dA_c(t)}{dt} &= R_{in}(t) - \left( \frac{C_L}{V_C} + \frac{Q_1}{V_C} + \frac{Q_2}{V_C}\right) \cdot A_c(t) + \frac{Q_1}{V_{P_1}} A_1(t)  + \frac{Q_2}{V_{P_2}} A_2(t),    \\
        \frac{dA_1(t)}{dt} &=  \frac{Q_1}{V_C} A_c(t) - \frac{Q_1}{V_{P_1}} A_1(t),  \\
        \frac{dA_2(t)}{dt} &=  \frac{Q_2}{V_C} A_c(t) - \frac{Q_2}{V_{P_2}} A_2(t),
    \end{align}
\end{subequations}
where the states $A_c(t)$, $A_1(t)$, and possibly $A_2(t)$ denote the amount of the drug in the central and peripheral compartments. 

By introducing the following reparameterisation into exchange and excretion parameters,
\begin{align*}
    k_{10} &= \frac{C_L}{V_C} & k_{12}&=\frac{Q_1}{V_C}  & k_{21} &= \frac{Q_1}{V_{P_1}} & k_{13} &= \frac{Q_2}{V_C} & k_{31} &= \frac{Q_2}{V_{P_2}} .
\end{align*}
we end up at the set of ODEs as described by Equations~\ref{eq:ode-model-2cmpt} and \ref{eq:ode-model-3cmpt}.

% \section{Appendix}\label{secA1}



\end{appendices}


\newpage
\bibliography{main}% common bib file

\end{document}